\documentstyle[graphicx]{elsart}
\newcommand\nn{\nonumber}
\newcommand\ba{\begin{eqnarray}}
\newcommand\ea{\end{eqnarray}}
\newcommand\alb{\begin{align}}
\newcommand\ale{\end{align}}
\newcommand\be{\begin{equation}}
\newcommand\ee{\end{equation}}
\newcommand{\br}[1]{\left( #1 \right)}
\newcommand{\brs}[1]{\left[ #1 \right]}

\newcommand{\brm}[1]{\left| #1 \right|}

\newcommand{\GeV}{\mbox{GeV}}
\newcommand{\MeV}{\mbox{MeV}}
\newcommand{\keV}{\mbox{keV}}

\newcommand{\vv}[1]{ {\vec #1}}

\begin{document}
\begin{frontmatter}

\def\figurename{Fig.}
\def\tablename{Tab.}



\title{Annihilation of $\bar p+p\to e^++e^-+ \pi^0$ and $\bar p+p\to \gamma + \pi^0$ through $\omega$-meson intermediate state}

\author{E.~A.~Kuraev,Yu.~M.~Bystritskiy,V.~V.~Bytev}
\address{\it Joint Institute for Nuclear Research-BLTP, 141980 Dubna, Moscow Region, Russia}
\author{A. Dbeyssi and E.~Tomasi-Gustafsson$^1$}
\address{
\it Univ Paris-Sud, CNRS/IN2P3, Institut de Physique Nucl\'eaire, UMR 8608, 91405 Orsay, France}
\thanks{Permanent address: CEA,IRFU,SPhN, Saclay, 91191 Gif/Yvette Cedex, France}

\date{\today}

\begin{abstract}
The $s$-channel annihilation of proton and antiproton into a neutral pion
and a real or virtual photon followed by lepton pair emission is studied. Such
mechanism is expected to play a role at  moderate values of the total
energy $\sqrt{s}$, when the pion is emitted around $90^{\circ}$
in the center of mass. A fair comparison with the existing data is obtained taking into account scattering and annihilation channels.
The cross section is calculated and numerical results are given in the kinematical range
accessible in the PANDA experiment at FAIR.
\end{abstract}
\end{frontmatter}
\maketitle


\section{Introduction}
\label{Introduction}
The antiproton-proton annihilation in various leptonic and hadronic channels will be
intensively studied by the PANDA collaboration \cite{Lutz:2009ff} at the complex
accelerator  FAIR (Darmstadt), where it is planned to accelerate
antiproton beams with momentum from 1.5 to 15 GeV/c and
average luminosity ${\cal L}= 1.6 \cdot 10^{32}$ cm$^{-2}$ s$^{-1}$
\cite{FAIR}. The $\bar p p$ annihilation into leptonic channels
is very challenging, as the hadronic background is higher by orders of
magnitude. In Refs. \cite{Re04,Sudol:2009vc} the annihilation reaction
$\bar p+p\to \gamma^* \to e^++e^-$ was investigated with the aim to extract of electromagnetic form factors of the proton in the time-like region. In this paper we focus on the annihilation reactions
\ba
&&\bar p+p\to \gamma^*+\pi^0 \to e^++e^-+ \pi^0,
\label{eq:eq1}\\
&&\bar p+p\to \gamma+\pi^0,
\label{eq:eq1a}
\ea
which contain also interesting information, in different aspects. Through the emission of one virtual photon between the proton--antiproton and the lepton pair, reaction (\ref{eq:eq1}) constitutes a unique opportunity to
determine time-like proton form factors in the 'unphysical' kinematical
region, $i.e.$, under the threshold of $\bar p+p$ annihilation into  a lepton pair
(i.e. for $0 < q^2 < 4M_p^2$, where $q^2$ is the momentum transferred to the proton)
\cite{Ad07}. Related processes are the pion electroproduction on the
proton $e^- + N \to e^- +  N  + \pi$, which is under intensive
experimental and theoretical study at MAMI and JLab \cite{Osipenko:2009jr} and the pion induced lepton pair production  $\pi + N \to N + \ell^- + \ell^+$, first investigated in \cite{Rekalo:1965}.  Pion scattering on nucleon and nuclei is part of the experimental program of the HADES collaboration at GSI \cite{Hades}.  The reaction (\ref{eq:eq1}) was studied in Ref.
\cite{Dubnickova:1995ns} near threshold. In Ref. \cite{Ad07} the calculation was extended introducing a larger set of diagrams and
studying the sensitivity to different parametrizations of proton FFs.
It was also suggested to extract information of the time-like nucleon axial
FF, through the production of a charged pion from the reaction
$\bar p+n \to e^++e^-+ \pi^-$. As no data exist for the nucleon axial
FF in the time-like region, predictions have been built on analytical extension of models which reproduce the data in the space-like region
\cite{Adamuscin:2007fk}.

The reaction (\ref{eq:eq1}) has also been investigated in backward
kinematics, in the region of large momentum transfer, with the
aim to get information of transition distribution amplitudes \cite{Pire:2005ax}. In Ref. \cite{Kroll:2005ni} the process $p\bar p\to \gamma \pi^0$ was considered in the frame of the Generalized Distributions Amplitudes approach. Neither the Regge behavior of the scattering amplitude nor the resonant character or annihilation amplitude was contained in those papers.

\begin{figure}
\begin{center}
\includegraphics[width=14cm]{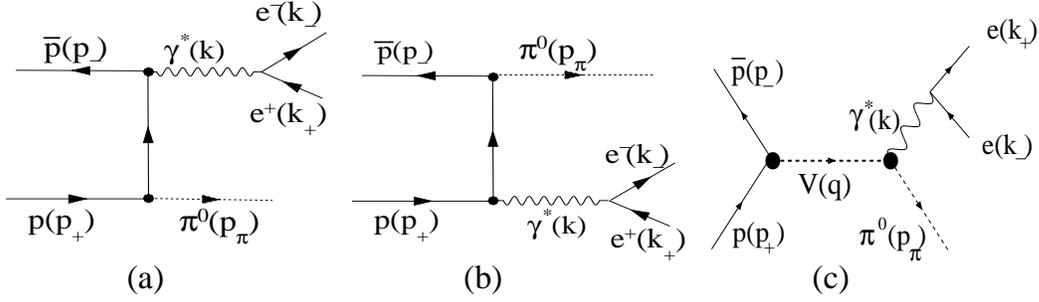}
\caption{Feynman diagram for $\bar p+p \to e^++e^-+ \pi^0$
(a) and (b) in $t$-channel ('scattering'); (c) in $s$-channel ('annihilation').}
\label{Fig:Fig1}
\end{center}
\end{figure}

The reaction mechanism which allows to access the electromagnetic structure of the proton is illustrated in Figs.~\ref{Fig:Fig1}a,b and will be denoted below as 'scattering' exchange channel. In forward (backward) kinematics the Regge character of exchanged nucleon reveals itself. In these kinematical conditions, the exchanged nucleon is close to mass shell. Therefore the vertex $\gamma^{(*)}*p\bar p$ can be described in terms of two phenomenological quantities $F_{1,2}$ which can be interpreted as proton form factors (FFs) which describe the partonic structure of the proton. This statement is valid in the kinematical region of near forward ($|t| \ll s$) or near
backward ($|u| \ll s$) scattering. In the region $|t|\sim|u| \sim s \gg M_p^2$
these amplitudes are
(unknown) functions of both kinematical variables. They have the form
$(M_p^2/s)^n\phi(t/s)$, where the exponent $n$ is determined by quark-counting
rules and are not related with the Dirac and Pauli
form factors of the proton. A similar behavior is expected for the vertex
of pion interaction with nucleons.

In collinear kinematics (near forward scattering and near
backward scattering) the amplitude of the scattering channel dominates, as the
 contribution of the annihilation channel is suppressed by the phase volume factor $|t|/s$ or $|u|/s$.  At emission angles near $90^{\circ}$, (center of mass reference frame implied) the cross section is dominated by the amplitude of the  the 'annihilation' vector meson exchange mechanism (see Fig. \ref{Fig:Fig1}c).  Intermediate states such as nucleonium ($p\bar p$ bound states), vector and scalar mesons (including radially excited meson states) can in principle contribute. In Ref. \cite{Ke95} the creation of narrow resonances was discussed. We focus our interest in the energy region outside the resonance production. Heavy vector meson intermediate states such as $\omega\br{1450}$ and $\omega\br{1650}$ mesons play an important role in limited kinematical range, where the total energy is close to their mass, due to Breit--Wigner character of relevant amplitudes the states with $\omega\br{1450}$ and $\omega\br{1650}$. Outside this region, they are suppressed by form factors since they are more extended objects.

In this paper, we focus on the processes (\ref{eq:eq1}), (\ref{eq:eq1a}), in the kinematical region where the $s$-channel is expected to dominate (large angle emission of the pion and moderate values of the total energy squared, $s$). The characterization of this mechanism (Fig. \ref{Fig:Fig1}c) is important not only to disentangle the information on proton form factors
(which are accessible through the 'scattering' mechanism) Fig. \ref{Fig:Fig1}a,b) but also to gain information on the
properties of the vector mesons. Here the parton (quark) structure of proton turns out to be essential. We will limit our considerations to $\omega$-exchange in the intermediate state. It is known from previous literature \cite{Witten:1983tw,Kaymakcalan:1983qq}
that the largest anomalous vertex is $\rho\omega\pi$, as it has the largest quark coupling.

The interaction of the vector $\omega$-meson with the nucleus in the vertex $\omega p p$, which  contains
information on the strong proton and meson couplings, can be
described in terms of proton vector form factors, in the time-like region of momentum transfer squared. They have, in principle, complex nature.  The vertex $\omega \to \pi\gamma^*$ can be described either by a phenomenological parametrization, or through a triangle vertex. Such triangle vertex, calculable in frame of Nambu-Jona-Lasinio model, is relevant also to the transition pion FF \cite{Volkov:1986zb,Volkov:2006vq,Radzhabov:2006ka}. Recently, large
interest has been raised in this field by the new data
from BaBar on pion FF at large momentum transfer \cite{Aubert:2009mc}.

The calculation will be compared to the unpolarized cross section and to the angular distribution for $\bar p+p\gamma+\pi$
annihilation, which was measured in the
region $2.911~\GeV \le\sqrt{s}\le 3.686~\GeV$ by the Fermilab E760 Collaboration \cite{Arm97}.

The plan of the present paper is as follows. In section 2 we calculate the
$\bar p+p\to \gamma+\pi^0$ process for the annihilation
channel: the kinematical variables are defined, the phase space and the matrix element of the reaction are calculated.
In section 3 the case of virtual photon is considered and  double differential cross section is calculated.
In section 4 a realistic parametrization for the vertices and the coupling constants are presented. In section 5 we compare the calculation to the data in condition of real photon emission around $90^{\circ}$, and predictions are given
for the double differential cross section for the reaction $\bar p+p\to \gamma^*+\pi^0$. In section 6 (Discussion and Conclusions) the domain of applicability of the present model is discussed, as well as its interference with the scattering channel exchange mechanism.

\section{Formalism for $\bar p+p\to \gamma + \pi^0$}

Let us consider first the process of proton-antiproton annihilation
into a real photon of momentum $k$ (with $k^2=0$) and a  neutral pion:
\ba
\bar p \br{p_-} + p\br{p_+} \to \gamma\br{k} + \pi^0\br{p_{\pi}}.
\label{eq:ProcessRealPhotonPi0}
\ea
Annihilation and scattering exchange channels play both a role, which will be discussed below. First the kinematics, which is common to both mechanisms, is discussed.

\subsection{Kinematics}
\begin{figure}
\begin{center}
\includegraphics[width=6cm]{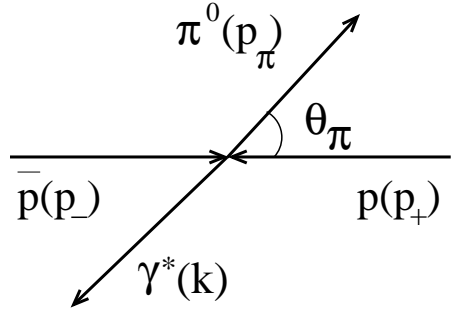}
\caption{Schematic view of the reaction $\bar p+p \to \gamma^*+ \pi^0$ in the center of mass system.}
\label{Fig:cms}
\end{center}
\end{figure}

The calculation is performed in the center of mass system (c.m.s.) (Fig. \ref{Fig:cms}).

Let us fix the momenta parametrization in c.m.s (in the form $p=\br{p_0,\vv{p}}$ where $\vv{p}$ is the 3-vector) as:
$$
    p_\pm = \br{E, \pm\vv{p}} = \br{\frac{\sqrt{s}}{2}, \,\pm\vv{p}},
    ~ p_\pm^2=M_p^2,~
    k = \br{k_0, \vv{k}},
    \nonumber
$$
$$
    q = p_+ + p_- = \br{2E, 0},~
    p_\pi = \br{E_\pi, -\vv{k}},~
    E=\frac{\sqrt{s}}{2},~ \beta=\frac{|\vec{p}|}{E}= \sqrt{1-4\displaystyle\frac{M_p^2}{s}},
   \nonumber
$$
where $E$($\beta$) is the energy (velocity) of the initial proton of mass $M_p$, $k_0$ is the  energy of the emitted photon and
$E_\pi$ is the pion energy. Then we can calculate the necessary scalar products:
\ba
    2\br{p_\pm q} = 4E^2 = s,\quad
    2\br{qk} = s-M_\pi^2,\quad
    2\br{p_\pm k} = \frac{s-M_\pi^2}{2}\br{1\pm \beta c_{\pi}},
\ea
where  and $c_{\pi}=\cos\br{-\vv{k},\vv{p}}$
is the cosine of angle between the momenta of the initial proton and the produced pion.

For convenience let us introduce the invariant variables:
\ba
    s &=& \br{p_++p_-}^2 = \br{k+p_\pi}^2=q^2,~
    \qquad
    u = \br{p_+-p_\pi}^2 = \br{p_--k}^2,
    \nn\\
    t &=& \br{p_--p_\pi}^2 = \br{p_+-k}^2,~
    \qquad\qquad
    s + t + u = 2M_p^2 + M_\pi^2, \nn
\ea
then
\ba
    &&2\br{p_+ p_-} = s-2M_p^2,~
    \qquad
    \qquad\quad
    2\br{p_- p_\pi} = -t+M_p^2+M_\pi^2,
    \nn\\
    &&2\br{p_+ p_\pi} = -u+M_p^2+M_\pi^2,~
    \qquad
    2\br{k p_\pi} = s-M_\pi^2,~
    \nn\\
    &&2\br{p_+ k} = -t+M_p^2,~
    \qquad
    \qquad\quad~
    2\br{p_- k} = -u+M_p^2.
    \nn
\ea
The phase volume of this reaction has the standard form:
\ba
     d\Phi_2     &\equiv&
    \br{2\pi}^4
    \frac{d\vv{k}}{\br{2\pi}^3 2k_0}
    \frac{d\vv{p_\pi}}{\br{2\pi}^3 2E_\pi}
    \delta^4\br{p_++p_--k-p_\pi}\\
    \nn
    &=&
    \frac{s-M_\pi^2}{2^5 \pi^2 s} d\Omega_\pi
    =
    \frac{s-M_\pi^2}{2^5 \pi^2 s} d\Omega_\gamma
\ea
where the final particles mass-shell $\delta$-functions restricts
the energies and the moduli of the momenta to:
\be
    E_\pi = \displaystyle\frac{s+M_\pi^2}{2\sqrt{s}},~
    \omega = \brm{\vv{k}} = \brm{\vv{p_\pi}} = \displaystyle\frac{s-M_\pi^2}{2\sqrt{s}},~
    \beta_\pi = \displaystyle\frac{\brm{\vv{p_\pi}}}{E_\pi} = \displaystyle\frac{s-M_\pi^2}{s+M_\pi^2},
\ee
and $\beta_\pi$ is the pion velocity. The cross section can be calculated using the following expression:
\ba
    d\sigma = \frac{1}{4 \cdot 4 I}
    \int \sum_{spins}\brm{{\cal M}}^2 d\Phi_2,
\label{eq:eq9}
\ea
where $I=\br{1/2}\sqrt{s\br{s-4M_p^2}}=s\beta/2$ is the invariant flux.

\subsection{$\bar p  + p\to \gamma + \pi^0$ annihilation through $\omega$ meson}

\begin{figure}
\begin{center}
\includegraphics[width=7cm]{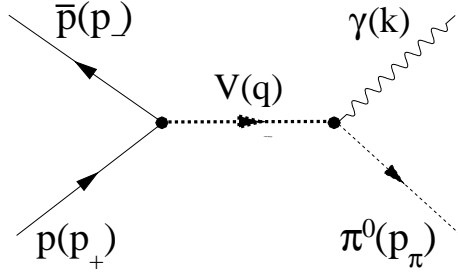}
\caption{Feynman diagram for the annihilation reaction $\bar p+p\to \gamma+\pi^0$, through vector meson exchange.}
\label{Fig:gamr}
\end{center}
\end{figure}

Let us consider the annihilation channel through vector meson exchange of the reaction
\be
\bar p(p_-)  + p(p_+)\to V(q)\to \gamma(k) + \pi^0(p_{\pi})
\label{eq:gamr}
\ee
as illustrated in Fig. \ref{Fig:gamr}.

The matrix element for the exchange of intermediate vector meson $V$ can be written as:
\ba
{\cal M}^{\gamma}= -
\frac{G_{V\pi\gamma}\br{q^2,0} G_{Vpp} \epsilon_{\mu\nu\rho\sigma} e^\nu q^{\rho}k^{\sigma}}{M_V[
s-M^2_{V}+i M_{V}\Gamma_{V}\br{q^2}]}
\br{g^{\mu\alpha}-\frac{q^\mu q^\alpha}{M_V^2}}
{\cal J}_{\alpha}^{\br{V}}\br{s},~
\label{eq:RealPhotonAnnihilationAmplitude}
\ea
where $V=\rho, ~\omega$,... $M_V$ and $\Gamma_V\br{q^2}$ are the mass and the
total width of intermediate vector meson $V$ with momentum $q$ ($q^2=s$),
$e^\mu$ is the polarization
vector of emitted real photon ($i.e.$, $\br{e k} = 0$).
In principle the total width of the intermediate vector meson depends on $q^2$. This dependence, which is function of the produced particle momentum, is particularly important in case of wide resonances. It is introduced to suppress unphysical contributions outside the resonance region \cite{Arbuzov:2010xi}. In the present case, as we are far from the resonance mass and we consider $\omega$ meson which has a small width, we neglect the $q^2$ dependence of the width
(i.e. $s \gg M_\omega^2, M_\omega \Gamma_\omega$).

The quantity $G_{V\pi\gamma}\br{q^2,k^2}$ describes the  $V\br{q}\to\pi\br{p_\pi}\gamma\br{k}$ vertex and it is given in section~\ref{SectionModels}. The current
related to the $p\bar p V$ vertex ${\cal J}_{\alpha}^{\br{V}}\br{q^2}$ has the form:
\ba
{\cal J}_{\alpha}^{\br{V}}\br{q^2}&=& \bar v\br{p_-} \Gamma_{\alpha}^{\br{V}}\br{q^2} u\br{  p_+},
\label{eq:VppCurrent}
\ea
where the vertex $\Gamma_{\alpha}^{\br{V}}$ is parameterized through two (strong) form factors $F_{1,2}^{\br{V}}$ in the following way:
\ba
\Gamma_{\alpha}^{\br{V}}&=&F_1^{\br{V}}\br{q^2}\gamma_\alpha +
\frac{i}{2M_p} F_2^{\br{V}}\br{q^2} \sigma_{\alpha\beta}q^\beta,
\label{eq:VppVertex}
\ea
where $\sigma^{\mu\nu} = \br{i/2}\br{\gamma^\mu\gamma^\nu-\gamma^\nu\gamma^\mu}$,
and the explicit expression for $F_{1,2}^{\br{V}}$ is given in section~\ref{SectionModels}.

The region of our interest is far outside the region of narrow
resonances which is defined as $|\sqrt{s}-M_R| \sim \Gamma_R \ll \sqrt{s}$,
where $M_R(\Gamma_R)$ is the mass(width) of the resonance.
In the region close to the resonance production the effects of formation
of bound states
is very important and can drastically change the value of the cross section
\cite{Ke95}. This effect appears in a very narrow range and it is not necessary to take it into
account in our case.

In principle, all vector mesons may contribute in the intermediate
state. Let us give the reasons why the contribution of an intermediate $\omega$-meson is dominant for this process, with respect to $\rho$-exchange. The value of the coupling $Vpp$ is
$g^2_{\omega pp}/(4\pi) =20\gg g^2_{\rho pp}/(4\pi)=0.55$. For the $V\pi\gamma$ vertex, the coupling of the vector meson with the quark is relevant. The coupling constant of the $\omega$ and $\rho$ mesons with the valence quarks in the proton or pion are quite different. The $\omega$--meson, being
isotopic singlet
constructively interacts with the quarks, unlike the $\rho$--meson which is
isovector and its interaction with quarks is destructive. That is the
reason for
$g_{\omega qq} \sim 3 g_{\rho qq}$. This is a large factor, as the square of this constant enters in the amplitude. The annihilation channel contains the vertex of interaction of two vector
mesons with the pion. This vertex has anomalous nature. Therefore, vector mesons should be of different species (for example if $\rho$ meson is in intermediate state,
then $\omega$-meson appears in the final state). As the decay width of $\rho$--meson is about ten times
larger that
the one for $\omega$--meson we conclude that the main mechanism of the
process under consideration
is:
\be
     p + \bar p \to \omega \to \pi + \rho
\ee
with the subsequent transition of the $\rho$--meson into a lepton pair or a real
photon \cite{PDG}.

About $\rho '$, $\omega '$ and other radial excitations of $\rho$ and $\omega$, an additional suppression is expected in the considered kinematical region, as form factors decrease faster, describing more extended objects.

Therefore, we consider only the contribution of the $\omega$ meson.  This simplifies also the analytical calculation, as  $F_2^\omega \ll F_1^\omega$ and can be neglected (see Ref. \cite{Machleidt:2000ge}).The vector current takes the simple form:
\be
{\cal J}_{\alpha}^{\br{\omega}}\br{q^2}= F_1^\omega \bar v\br{p_-} \gamma_{\alpha} u\br{p_+}.
\label{eq:VppCurrent1}
\ee
Using the relation
\ba
&&\epsilon_{\mu\alpha\rho\sigma}
\epsilon_{\nu\alpha\gamma\delta}
q^\rho k^\sigma q^\gamma k^\delta
\sum_{spin} {\cal J}^\mu {\cal J}^{*\nu}
=2 s^2 \vec k^2
\br{2-\beta^2 \sin^2\theta_\pi} \brm{F_1^\omega}^2,
\label{eq:LeptonSpinsSum}
\ea
we obtain for the differential cross section
\ba
    \br{\frac{d\sigma}{d\Omega_\gamma}}^{ann}
    &=&\sigma_0(q^2) \br{2-\beta^2 \sin^2\theta_\pi},\nn\\
    \sigma_0(q^2)&=&\frac{\brm{G_{\omega\pi\gamma}\br{q^2,0}}^2 \br{q^2-M_\pi^2}^3 g_{\omega pp}^2} {2^9 \pi^2 q^2 \beta \brm{q^2-M^2_{\omega}+i M_{\omega}\Gamma_{\omega}}^2 M_\omega^2}\brm{F_1^\omega}^2,
\label{eq:dsigma}
\ea
where the $q^2=s$ dependence is contained in $\sigma_0$, which has the dimension of a cross section. The total cross section as function of $q^2$ is:
\be
    \sigma_{ann}(q^2)= 8\pi \sigma_0(q^2) \left( 1-\frac{\beta^2}{3}\right ).
    \label{eq:sigma}
\ee

The approximation of $\omega$-meson in the intermediate state is well beyond the precision on the final cross section, that is evaluated to be $\approx 10\%$. This estimation takes into account the accuracy of $SU(3)$ symmetry and the contributions of the neglected mesons.

\subsection{$t$ and $u$ channel exchange for $\bar p  + p\to \gamma + \pi^0$ }
\begin{figure}
\begin{center}
\includegraphics[width=12cm]{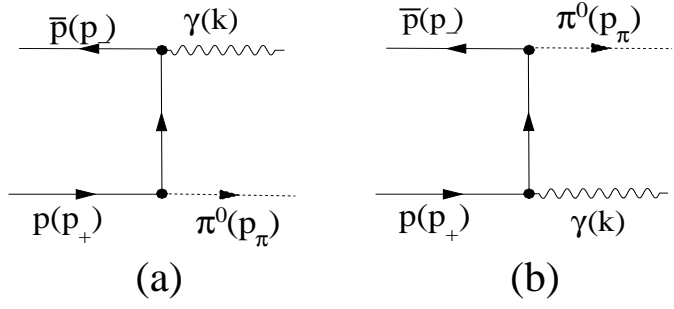}
\caption{Feynman diagram for the reaction $\bar p+p\to \gamma+\pi^0$, (a) in $t$ and (b) in $u$ channels.}
\label{Fig:gamut}
\end{center}
\end{figure}
Let us consider the diagrams illustrated in Fig.
\ref{Fig:gamut}. The corresponding matrix element can be written as:
\be
    {\cal M}_{sc}=
        e g_{\pi NN}\bar v\br{p_-} \left [ \gamma_5 \displaystyle\frac{\hat q_1+M}{t-M^2} \Gamma_\mu + \Gamma_\mu \displaystyle\frac{ \hat q_2+M}{u-M^2} \gamma_5 \right ] u\br{p_+}
        e^\mu,
\ee
where $e$ is the proton charge, $g_{\pi NN}$ is the pseudoscalar coupling constant of the pion with the nucleon (the pseudovector coupling has momentum suppression \cite{Gasiorowicz}), $q_1 = p_+ - k$, $q_2 = p_+ - p_\pi$ and the proton electromagnetic vertex $\Gamma_\mu\equiv\Gamma_\mu\br{k^2=0}$ is parameterized in terms of form factors $F_{1,2}$:
\ba
\Gamma_\mu\br{k^2} &=&F_1\br{k^2}\gamma_\mu +
\frac{i}{2M_p} F_2\br{k^2} \sigma_{\mu\nu}k^\nu,
\label{eq:EMVertex}
\ea
which, in the real photon limit, coincide with the static values:
$F_{1}$=1, $F_{2}$=1.798. The differential cross section is:
\be
d\sigma_{sc}=\displaystyle\frac{1}{8s}\sum|{\cal M}_{sc}|^2d\Gamma_2
   \label{eq:eqms}
\ee
where the phase space can be written as:
\ba
d\Gamma_2&=&(2\pi)^{-2}\displaystyle\frac{d^3k}{2k_0}\displaystyle\frac {d^4q}{\sqrt{s}}\delta(q^2-m_{\pi}^2)
=\displaystyle\frac{1}{4\pi}k_0 dk_0 d\cos\theta
\delta [(p_++p_--k)^2-M^2]\nn\\
&=&
\displaystyle\frac{s-m_{\pi}^2}{16\pi s}d\cos\theta.
   \label{eq:msu}
\ea
We used the relation $(p_++p_--k)^2-M^2=s-2\sqrt{s}k_0-m_{\pi}^2$.

The differential cross section for $t$ and $u$ channel, and their interference  becomes:
\be
d\sigma=\displaystyle\frac{1}{2s}
\displaystyle\frac{s-m_{\pi}^2}{16\pi s}  \sum|{\cal M}_{sc}|^2d\cos\theta,
   \label{eq:s2}
\ee
with
\ba
    \sum|{\cal M}_{sc}|^2&=&
    e^2 g_{\pi NN}^2\br{
        -\frac{s}{t-M^2} F_t R_t
        -\frac{s}{u-M^2} F_u R_u
        +\frac{s}{2M^2} F_2^2\br{s}
    },
    \\
    F_t &=& F_1^2\br{t}+2F_1\br{t}F_2\br{t}+\displaystyle\frac{1}{2}F_2^2\br{t},
    \nn\\
    F_u &=& F_1^2\br{u}+2F_1\br{u}F_2\br{u}+\frac{1}{2}F_2^2\br{u}.
    \nn
%
	\label{eq:m2ut}
\ea
The effects of strong interaction in the initial state interaction
which comes from exchange by vector and (pseudo) scalar
mesons between proton-antiproton are really essential here, in the
scattering channel. They effectively lead to the Regge form of the amplitude of the scattering channel. Thus the $t$ and $u$ diagrams are suppressed by adding a general Regge factor $R_t$, $R_u$, in the following form:
\be
  R(t)= \left (\displaystyle\frac{s}{s_0}\right )^{2[\alpha(t)-1]},~
    \alpha_p(t) =  \displaystyle\frac{1}{2}+r\displaystyle\frac{\alpha_s}{\pi}\displaystyle\frac{t-M^2}{M^2}
    \label{eq:eqrt}
\ee
\be
  R(u)= \left (\displaystyle\frac{s}{s_0}\right )^{2[\alpha(u)-1]},~
    \alpha_p(u) =\displaystyle\frac{1}{2}+r\displaystyle\frac{\alpha_s}{\pi}\displaystyle\frac{u-M^2}{M^2}
   \label{eq:eqru}
   \ee
where $r\alpha_s/\pi\simeq 0.7$ and $s_0\simeq 1$ GeV$^2$ \cite{Kaidalov:2001db} can be considered fitting parameters.

This Regge form of amplitude incorporates in principle infinite number of resonances,
(i.e. $\Delta\br{1232}$ and others). As for excited resonances like $N^*\br{1440}$ they
belong to daughter Regge trajectory which contribute power suppressed value compared to the leading one.
This is also included in the estimation of the claimed 10\% error.

\section{Formalism for the reaction $\bar p+p\to e^++e^- + \pi^0$}
Let us consider the annihilation channel of the reaction:
\ba
\bar p\br{p_-} + p\br{p_+}
\to
\gamma^*\br{k} + \pi^0\br{p_\pi}
\to
e^+\br{k_+} + e^-\br{k_-} + \pi^0\br{p_\pi}
\label{eq:PairProcessWithMomenta}
\ea
where the four momenta of the particles are indicated in parenthesis.

We focus here on the mechanism of annihilation through $\omega$ meson exchange. The general expression of the matrix element, corresponding to
the diagram of Fig. \ref{Fig:Fig1}c, for the exchange of a vector meson $V$, is:
\ba
{\cal M}^\pm=4\pi\alpha\frac{G_{V\pi\gamma^*}(q^2,k^2) g_{Vpp}}{e}
\frac{\epsilon_{\mu\nu\rho\sigma }q^{\rho}k^{\sigma}}
{k^2M_{V}\br{q^2-M^2_{V}+iM_{V}\Gamma_{V}}}{\cal J}_p^{\mu}{\cal J}_e^{\nu},
\label{eq:Matrix}
\ea
where $e=\sqrt{4\pi \alpha}$ is the elementary electric charge
($\alpha \approx 1/137$). The lepton electromagnetic current has a standard form
\ba
{\cal J}_e^{\nu}=\bar u(k_-)\gamma^{\nu}v(k_+),
\label{eq:LeptonCurrent}
\ea
and the electromagnetic current related to the $p\bar p V$ vertex is
\be
{\cal J}_p^{\mu}=\bar v(p_-) \Gamma_\omega^{\mu} u(p_+), \qquad
\Gamma_\omega^{\mu}=F_1^\omega\br{q^2}\gamma^{\mu}.
\label{eq:ProtonCurrent}
\ee
The currents obey gauge invariance:
$k_\nu {\cal J}_e^{\nu} = q_\mu {\cal J}_p^{\mu} = 0$.
Again, only $\omega$ exchange is considered and $F_2$ is neglected.

The cross section of the process (\ref{eq:PairProcessWithMomenta})
can be written in the standard form
\ba
    d\sigma = \frac{1}{4I} \int \sum_{spin} \brm{\cal M}^2 d\Phi_3,
    \label{eq:GeneralCrossSection}
\ea
where $d\Phi_3$ is the phase volume of the process:
\ba
    d\Phi_3
    &=&
    \br{2\pi}^4 \delta^4\br{p_+ + p_- - k_+ - k_- - p_\pi}
    \frac{d^3 \vec k_+}{\br{2\pi}^3 2E_+}
    \frac{d^3 \vec k_-}{\br{2\pi}^3 2E_-}
    \frac{d^3 \vec p_\pi}{\br{2\pi}^3 2E_\pi}
    =\nn\\
    &=&
    \frac{1}{\br{2\pi}^4}
    \beta_\pi E_\pi dE_\pi dc_\pi\frac{1}{2} d\Phi_e,
    \nn\\
    d\Phi_e &=&
    \frac{d^3 \vec k_+}{2E_+}
    \frac{d^3 \vec k_-}{2E_-}
    \delta^4\br{k - k_+ - k_-},
\ea
where $E_\pm$ and $E_\pi$ are the energies of the electron, the positron and the
pion in the final state.

Since our aim is to calculate the angular distribution and energy spectrum
of the final pion, we insert the unit integration
\ba
    \int d^4 k ~ \delta^4\br{k-k_+-k_-} = 1.
\ea
Performing the integration over the final leptons momenta:
\ba
&&\int
d\Phi_e
\sum_{spin} {\cal J}_e^\mu {\cal J}_e^{*\nu}
=-\frac{2\pi}{3}k^2 \br{g^{\mu\nu} - \frac{k^\mu k^\nu}{k^2}} \phi\br{k^2},
\ea
where
\ba
\phi\br{k^2} = \br{1+\frac{2m_e^2}{k^2}}\sqrt{1-\frac{4m_e^2}{k^2}},
\ea
one obtains the following expression for the double differential cross section:
\ba
    d^2\sigma^\pm &=& \sigma_0^\pm dW, \nn\\
    \sigma_0^\pm(q^2)
    &=&
    \frac{\alpha \, (q^2)^2 \brm{G_{\omega\pi\gamma^*}\br{q^2,k^2}}^2
        \brm{F_1^\omega\br{q^2}}^2 g_{\omega pp}^2
    }
    {
     \beta~M_\omega^2~
        \brs{\br{q^2-M^2_\omega}^2+M_\omega^2\Gamma_\omega^2}
    },
    \nn\\
    dW &=&
    \frac{\beta_\pi^3 E^3_{\pi}\br{2-\beta^2\sin^2\theta_{\pi}}}
    {48 s \pi^2 k^2} \phi\br{k^2}dE_{\pi}dc_{\pi},
    \label{eq:eq2a}
    \\
    k^2 &=& s+M_\pi^2-2E_{\pi}\sqrt{s} > 4m_e^2, \nn
\ea
where we put $k^2=0$ in the vertex $\omega\pi\gamma^*$ since we consider the case
of small virtuality.
The energy of the final pion distribution can be obtained by the integration
over the angle:
\ba
\frac{d\sigma^\pm}{dE_{\pi}}
&=&
\int\limits_{-1}^1 dc_\pi
\frac{d^2\sigma^\pm}{dE_{\pi}dc_{\pi}}
=
\sigma_0^\pm(q^2)
\frac{\beta_\pi^3 E^3_{\pi}}
    {12 \, q^2 \pi^2 \, k^2} \phi\br{k^2}
\br{1-\displaystyle\frac{\beta^2}{3}},\nn\\
\beta_\pi E_\pi&=&\sqrt{E_\pi^2-M_\pi^2}.
\label{eq:EnergySpectrum}
\ea
\section{Models of Form Factors and Vertices}
\label{SectionModels}

In principle, one should include initial state interaction (ISI). ISI strongly modifies the cross section, at near threshold energies. It is evaluated introducing a multiplicative factor $C$, the Coulomb factor, which takes into account multiphoton exchange:
\be
C=\displaystyle\frac{x}{e^x-1},~x=\displaystyle\frac{\alpha\pi}{v},
\ee
where $v$ is the incident relative velocity in $c$ units. In the high energy range considered here, this factor is close to unity. In our case, this factor increases the cross section, as the interaction occurs between opposite charges. For example, at $\sqrt{s}=2$ GeV, $v=0.34~c$, $C=1.6$. As far as meson exchanges are concerned, let us remind that in $\bar p p $ annihilation, one pion exchange is suppressed by the factor $Mp^2/s$ which becomes more and more important at high energies,the $\omega$ meson is included in the $\omega NN$ form factor and heavier vector meson are suppressed for the reasons given above.

Following \cite{FernandezRamirez:2005iv}, the form factor  $F_{1,2}^{\omega}\br{q^2}$ can be parametrized as:
\be
    F_1^{\omega}(q^2) = \frac{\Lambda_{\omega}^4}{\Lambda_{\omega}^4+(q^2-M_{\omega}^2)^2},
 \label{eq:FFVpp}
 \ee
with normalization $F_1^{\omega}(M_{\omega}^2) = 1$.  $\Lambda_{\omega}= 1.25$  GeV  is an empirical cut-off.  We took $M_{\omega}=782.65\pm 0.19$ MeV.

This parametrization has been built for the space-like region of momentum transfer. In the time-like region, one should consider an extension of this formula which obeys analyticity, in particular introducing an imaginary part.
This procedure has been suggested in literature for electromagnetic form factors of the nucleon in Ref. \cite{TomasiGustafsson:2005kc}, for $\rho$-meson \cite{Adamuscin:2007dt} and for $a_1$ meson in Ref. \cite{TomasiGustafsson:2008kn}. Such parametrizations are very reliable when data exist to constrain the parameters. This is not the present case, and such procedure would add more  uncertainties.

\begin{figure}[h]
\begin{center}
\includegraphics[width=6cm]{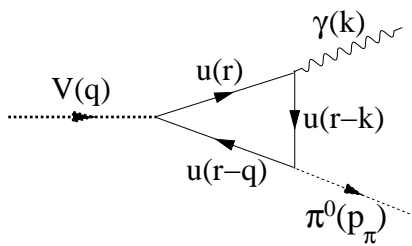}
\caption{Triangle diagram for $V\to\pi\gamma$.}
\label{Fig:trianglevertex}
\end{center}
\end{figure}

The quantity $G_{\omega\pi\gamma^*}$ is related to the matrix element of the transition  $\omega(q)\to\pi(p)\gamma(k)$  (Fig. \ref{Fig:trianglevertex}) and it is in principle momentum dependent:
\ba
M(\omega\to\pi\gamma)=\frac{G_{\omega\pi\gamma^*}}{M_{\omega}}\br{q^2,k^2}\epsilon_{\mu\nu\alpha\beta}q^{\alpha}k^{\beta}
\epsilon_{\rho}^{\mu}\epsilon_{\gamma}^{\nu},
\ea
where we introduced the polarization vectors of the photon and of the vector meson.

It can be calculated in the following \cite{VolkovPervushin}:
\be
    \brm{G_{\omega\pi\gamma}\br{q^2,k^2}}^2 =
    9\frac{\alpha}{\pi^3}
    \frac{g_{\omega uu}^2 M_\omega^2}{F_\pi^2} \brm{I\br{q^2,k^2}}^2,
 \label{eq:Vertex}
\ee
where we used the Goldberger--Treiman relation $g_{\pi uu} / m_u = 1/F_\pi$,
and $F_\pi = 93~\MeV$ \cite{PDG} is the pion decay constant.
The constants $g_{\omega uu}$ and $g_{\pi uu}$
correspond to the coupling of $\omega$ and $\pi$ mesons with the
light $u$-quarks in the loop.
We take the following values \cite{Volkov:2009zzb}:
$ g_{\omega uu} = 5.94$ and $g_{\pi uu} = 2.9$.

$I\br{q^2,k^2}$ is the internal quark loop integral which will
be derived in the Appendix~\ref{AppendixLoopIntegral}. Due to the requirement of absence of real quark in the intermediate state, the quantity $I\br{q^2,k^2}$ must be real:
\ba
    I\br{q^2,k^2} = \frac{m_u^2}{2\br{k^2-s}}
    \brs{
        \ln^2\br{\frac{q^2}{m_u^2}}-\ln^2\br{\frac{k^2}{m_u^2}}
    },~
    k^2 \gg m_u^2,~q^2\gg m_u^2,
\ea
where $m_u$ is the constituent quark mass.
We verified that our results for real and virtual photons agree in the limit $k^2 \to m_u^2$:
\ba
    I\br{s,k^2\to m_u^2} \approx -\frac{m_u^2}{2s} \ln^2\br{\frac{s}{m_u^2}}.
\ea
In order to check the consistency of the values of the parameters, we evaluate
the width of the radiative decay
$\omega \to \pi\gamma$:
\ba
\Gamma\br{\omega\to\pi^0\gamma}
=
\frac{\alpha}{192}
\frac{M_\omega^3}{F_\pi^2}
\frac{g_{\omega uu}^2}{\pi^4}
\br{1 - \frac{M_\pi^2}{M_\omega^2}}^3
\approx
550 ~\keV.
\ea
Thus the decay branching is equal to:
\ba
BR\br{\omega \to \pi^0\gamma}
=
\frac{\Gamma\br{\omega\to \pi^0\gamma}}{\Gamma_\omega}
=
6.5\%,
\ea
which is in a fair agreement with the value:
$BR^{exp.}\br{\omega \to \pi^0\gamma}=\br{8.28\pm 0.28}\%$ \cite{PDG}.
The values of the constants used in the calculation are   $\Gamma_{\omega}=8.49\pm 0.08$ MeV, $Br(\omega\to \pi\gamma)=(8.28\pm 0.28)\cdot 10^{-2}$, $\Gamma_ {\omega\to\pi\gamma} =  0.70297$ MeV,  $\displaystyle\frac{g^2_{\omega NN}}{4\pi}=20$.

Alternatively, one can choose a phenomenological parametrization for the vertex
$\omega\to\pi\gamma^{(*)}$, based on monopole dependencies on $q^2$ and $k^2$:
\ba
G_{\omega\pi\gamma^*}\br{q^2,k^2}= \frac{G_{\omega\pi\gamma}\br{0,0}}{(1+q^2/M_\omega^2)(1+k^2/M_\omega^2)},
\label{eq:eqv}
\ea
where, the constant $G_{\omega\pi\gamma}\br{0,0}$ is derived from the radiative decay
$\omega\to\pi\gamma$ \cite{Rekalo:2000su}:
\ba
\Gamma\br{\omega\to\pi\gamma}
=
\frac{M_\omega\alpha}{24}|G_{\omega\pi\gamma}(0,0)|^2\left (
1 - \frac{M_\pi^2}{M_V^2}\right )^3.
\label{eq:eqv1}
\ea
%

\section{Results}

\begin{figure}
\begin{center}
\includegraphics[width=9cm]{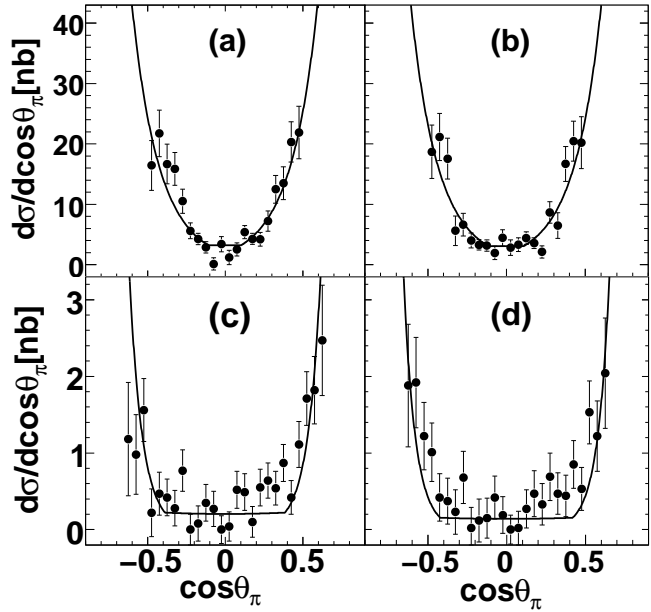}
\caption{ Angular distributions for different values of the c.m.s. energy: a) $\sqrt{s}=2.975$ GeV, (b) $\sqrt{s}=2.985$ GeV, (c) $\sqrt{s}=3.591$ GeV, (d) $\sqrt{s}=3.686$ GeV. The data are from \protect\cite{Arm97}, the line is the result of the model (see text).}
\label{Fig:ang}
\end{center}
\end{figure}

No experimental data exist on the reaction $\bar p +p\to e^++e^-+\pi^0$, but the reaction $\bar p +p\to\gamma+\pi^0$ was measured in the
region $2.911~\GeV\le\sqrt{s}\le 3.686~\GeV$ by the Fermilab E760 Collaboration \cite{Arm97} and data exist on the cross section and the angular distribution. It is in principle possible to adapt the present model to $\gamma$ production, and compare the calculation to the data in the angular region around $90^{\circ}$, where the present mechanism is expected to be dominant (Fig \ref{Fig:ang}).

\begin{figure}
\begin{center}
\includegraphics[width=10cm]{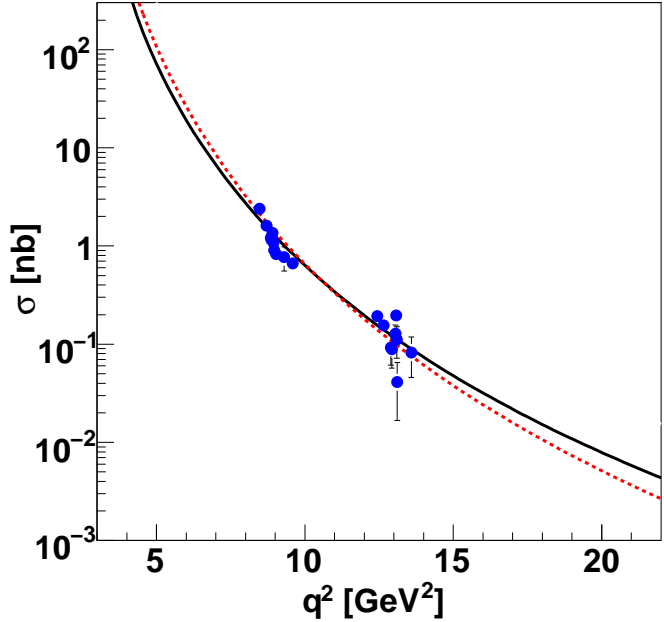}
\caption{ (Color online) $q^2$ dependence of the cross section. The data and the calculations are  integrated for $|cos\theta_{\pi}|<0.2$  for two choices of the $G_{\omega\pi\gamma}$ coupling: from the parameterizations (\ref{eq:Vertex}) (red, dashed line) and from monopole dependence (black, solid line).
}
\label{Fig:csreal}
\end{center}
\end{figure}

A good description with the existing data can be obtained, for parametrization (\ref{eq:Vertex}) of the coupling constant $g_{V_\pi\gamma}$, taking into account an additional coupling factor of
\be
    F_{1q}^{\omega}(q^2) = \left [\frac{\Lambda_{\omega}^4}{\Lambda_{\omega}^4+(q^2-M_{\omega}^2)^2}\right ]^{1/3},
 \label{eq:FFVqq}
 \ee
 due to the interaction of the vector meson with the constituent quarks in the vertex $Vu\bar u$. A good agreement with the data is also obtained using the phenomenological parametrization
(\ref{eq:eqv}) and $\Lambda_{\omega}=1.25$ GeV. This is shown in Fig. \ref{Fig:csreal}, where the results on the cross section integrated in the range $|\cos\theta_{\pi}|<0.2$ are reported and compared with the present calculation in $s$-channel.

In Fig. \ref{Fig:ang} the angular
distributions are shown for four values of the total c.m.s. energy: (a) $\sqrt{s}=2.975$ GeV, (b) $\sqrt{s}=2.985$ GeV, (c)
$\sqrt{s}=3.591$ GeV, (d) $\sqrt{s}=3.686$ GeV, together with the results of the calculation.  A good agreement for the $s$-channel calculation, Eq. (\ref{eq:dsigma}), is obtained in the angular region around $\cos\theta_{\pi}\approx 0$, because, as expected, this mechanism is applicable at these angles.

In order to reproduce the data in the full range where they are available, we have to consider also $t$ and $u$ exchanges. Here we focus to $s$-channel properties and we apply a simplified procedure, where we apply the $s$, $t$ and $u$ channel in separated kinematical regions, hence neglecting their interference.

From the data it appears that the angular distribution flattens in a range around $\cos\theta\simeq 0$, which depends on $s$. In this central region we apply the $s-$channel result, Eq. (\ref{eq:dsigma}). In the forward and backward regions, we deconvolve in Eq. (\ref{eq:m2ut}) the $u$ and $t$ contributions and apply the suppression due to Regge factors.
The results are shown in \ref{Fig:ang}.

For the $s$-channel annihilation, which is the object of the present paper, the differential and total cross sections have been numerically calculated for the reaction $\bar p +p\to e^++e^-+\pi^0$ and the results are illustrated for $q^2=7$ GeV$^2$ and for the parametrization (\ref{eq:Vertex}) of the $\omega\pi\gamma^*$ vertex.

\begin{figure}
\begin{center}
\includegraphics[width=10cm]{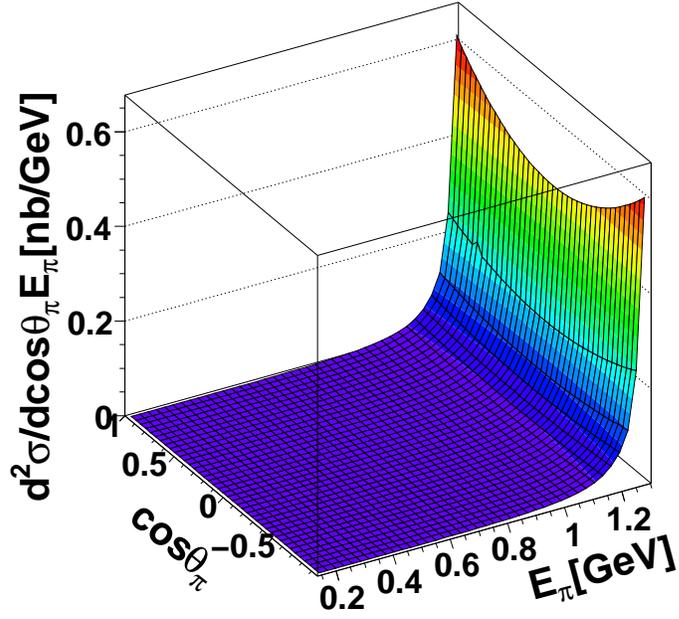}
\caption{(Color online) Bi-dimensional plot of the double differential cross section  for the process $\bar p+p\to \gamma^*+\pi^0$, as function of $E_\pi$ and $\cos\theta_{\pi}$ at  $q^2=7$ GeV$^2$.}
\label{Fig:bidim}
\end{center}
\end{figure}

The bidimensional plot of the cross section as a function of the c.m.s. angle and energy of the pion is shown in \ref{Fig:bidim}. The angular and energy dependences are fixed by the model. The absolute value of the cross section strongly depends on the cutoff parameter taken for the description of the $\omega NN$ vertex, which was taken here as for the real photon data.

In case of virtual photon, the $G_{\omega\pi\gamma}$ coupling depends on $q^2$ and $k^2$. The phenomenological parametrization can be modified adding a monopole dependence on $k^2$. For the parametrization (\ref{eq:Vertex}), the triangle integral
has in general more complicated form \cite{Ametller:1983ec}. Neglecting the pion mass, one can use the symmetry properties of the triangle loop (see Appendix).

\begin{figure}
\begin{center}
\includegraphics[width=10cm]{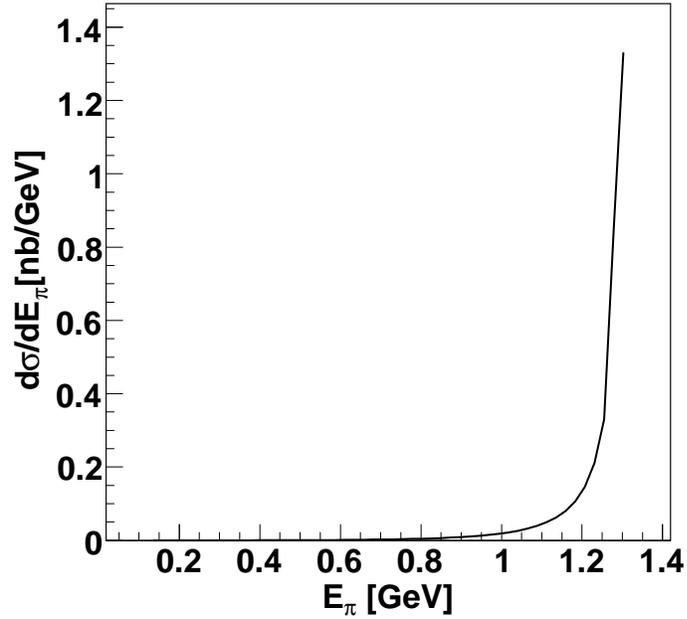}
\caption{Differential cross section for the process $\bar p+p\to \gamma^*+\pi^0$, as a function of $E_\pi$ at $q^2=7$ GeV$^2$.
}
\label{Fig:Epi}
\end{center}
\end{figure}

The bidimensional cross section as a function of the pion variables, at fixed total energy $s=$ 7 GeV$^2$ is shown in Fig.  \ref{Fig:bidim} and the projection on the pion energy is drawn in Fig. \ref{Fig:Epi} for the parameterization (\ref{eq:Vertex}) of the  $G_{V\pi\gamma}$ vertex. One can see that the cross section has a smooth angular dependence and it is larger when the pion energy is larger.

\section{Discussion and Conclusions}

A hadronic model for the annihilation channel in the processes $p+\bar p \to \pi^0+  \gamma\br{\gamma^*}$ is presented. This reactions is described by two classes of diagrams. We focused here on the annihilation type diagrams. The scattering type diagrams with a nucleon in the intermediate state were considered earlier, in Ref. \cite{Ad07} and Refs. therein.

We considered $\omega$ meson intermediate state as the main contribution because it contains the same $u$ and $d$ quarks as the final $\pi^0$, excluding the $\rho$-meson due to the small $\rho NN$  coupling as well as higher resonances. We applied our calculation in the energy region outside the resonance production. We compared our calculation to the existing data on $p\bar p \to \pi^0 \gamma$, and found a good description taking into account also $t$ and $u$ channel, in forward and backward angular regions. The formulas obtained in this paper describe the annihilation channel through $\omega$--meson (which gives an almost isotropic contribution in the center of mass system) as well as the near forward or near backwards kinematics (Reggeized nucleon in $t$ and $u$ channels). We reproduce the experimental evidence that the range of the central angular region, where $s$-channel dominates, increases when the energy increases.

As it was shown in Ref. \cite{Ad07}, the contribution of the $t$-channel can be parameterized in terms of two unknown functions $F_{1,2}\br{t,u}$ which depend of both variables $t$ and $u$ ($s+t+u=2M^2$) and rapidly decrease
in the region $s\sim\brm{t}\sim\brm{u}$ as $F_{1,2}\br{t,u}\sim \br{M^2/s}^n \ll 1$ where $n > 1$.
Let us note that in the limits $\brm{t} \ll s$ or $\brm{u} \ll s$ these quantities do coincide with the electromagnetic form factors of the proton. In these kinematical regions the
Regge--factors have to be included into the differential cross section. These factors rather strongly suppress the cross section.

Initial state interaction through meson exchange is effectively taken into account in our calculation, in the following way:
- in the annihilation channel it is in principle included in the form factors
which describe the vertex $\bar pp\omega$, which is in principle complex.
We should note that in model which we use and which was described in \cite{FernandezRamirez:2005iv} this coupling is real. In general it is possible to use parametrizations built for the space-like region and make an analytical extension to time-like region. However any extension of the parametrization requires extra parameters, which can not be presently constrained, and induce unnecessary complications in the model.
- in the scattering channel the effects of ISI are included in the Regge factor of the amplitudes.

Let us note that in case of $\gamma^*$ production, with subsequent conversion to electron positron pair, in the annihilation channel, one can take into account an intermediate production of $\rho$ meson form the $\omega$ by the following replacement:
\be
\displaystyle\frac{e^2}{k^2}\to \displaystyle\frac{G_{\rho e^+ e^-}G_{\rho u u }} {k^2-M_\rho^2+iM_\rho\Gamma\rho}
\ee
where $G_{\rho u u }=G_{\rho pp }$ and $ G_{\rho e^+ e^-}$ is   determined from the width of the decay $\rho\to e^++e^-$:
\be
G_{\rho e^+ e^-}^2= \displaystyle\frac{12\pi\Gamma_{\rho\to e^+e^-}}{M_{\rho}},
\ee
with $\Gamma_{\rho e^+ e^-} = 7 \cdot 10^{-3}$ MeV \cite{PDG}.

We estimate an accuracy of our calculation of ~$10 \%$ which is evaluated from the precision of Nambu--Jona-Lasinio model and the contributions of other vector mesons which we do not consider here.

The present approach can be generalized to all pseudoscalar mesons,
$\pi$, $\eta$, $\eta\prime$...

\section{Acknowledgments}

One of us (A.D.) acknowledges the Libanese CNRS for financial support. The authors are grateful to Dr. G. I. Gakh, J. Van de Wiele and S. Ong for interesting discussions and remarks. This work was done in frame of JINR-IN2P3 collaboration agreement and of GDR n.3034 'Physique du Nucl\'eon' (France) and partly supported by the grants RFBR 10-02-01295 and Bielorussia-JINR 2010.

\appendix


\section{Loop integral}
\label{AppendixLoopIntegral}

In (\ref{eq:Vertex}), the vertex $V\br{q} \to \pi\br{p_\pi} \gamma\br{k}$
was expressed in terms of quark loop integral:
\ba
    &&I\br{q^2,k^2}
    =\nn\\
    &&-
    \int
    \frac{d^4 p_q}{i \pi^2}
    \frac{m_u^2}
    {
        \br{p_q^2-m_u^2 +i0}\br{\br{p_q-q}^2-m_u^2+i0}\br{\br{p_q-k}^2-m_u^2+i0}
    }
    \nn\\
    &&=
    \int\limits_0^1 dx
    \int\limits_0^{1-x} dy
    \frac{1}{1 - \frac{q^2}{m_u^2} x y - \frac{k^2}{m_u^2} x z -\frac{M_\pi^2}{m_u^2} y z - i0}
    ,
    \label{eq:iqk}
\ea
where $m_u$ is the mass of $u$-quark which circulates in the loop.
The value of this integral in case of real photon (i.e. $k^2=0$)
is well-known (see, for instance, Ref. \cite{Gunion:1987ke}):
\ba
    I\br{q^2,0} = I_2\br{\frac{q^2}{m_u^2}, \frac{M_\pi^2}{m_u^2}},
\ea
where $I_2\br{a,b}$ was defined in \cite{Gunion:1987ke}:
\ba
    I_2(a,b) &=&
    \frac{2}{a-b}\brs{f\br{\frac{1}{b}} - f\br{\frac{1}{a}}},
    \\
    f(x) &=& \left\{
        \begin{array}{ll}
            -\brs{ \arcsin\br{\frac{1}{2\sqrt{x}}} }^2, & x > \frac{1}{4} \\
            \frac{1}{4} \brs{ \log\br{\frac{\eta_+}{\eta_-}} - i\pi }^2, & x < \frac{1}{4}
        \end{array}
        \right.,
\label{eq:I2ab}
\ea
and $\eta_\pm=\frac{1}{2}(1\pm\sqrt{1-4x})$.

Applying this formula in the triangle Feynman diagram  to the quark loop in the intermediate state, one should drop the imaginary term $i\pi$ (Fig. \ref{Fig:trianglevertex}). Following confinement properties, real quarks on mass shell are not allowed.

In the process involving a virtual photon, (i.e., $k>0$), this integral
has a more complicated form. However one can find a simple expression neglecting the pion mass. Indeed, the expression (\ref{eq:iqk}) is symmetric with respect to $q^2$, $k^2$, and $M_{\pi}^2$, therefore, one can use Eq. (\ref{eq:I2ab}) with
$$ I\br{q^2,k^2}=I_2\br{\frac{q^2}{m_u^2}, \frac{k^2}{m_u^2}}.$$



\end{document}